\documentclass{aa}  

\usepackage{graphicx}
\usepackage[colorlinks]{hyperref}
\usepackage{afterpage}
\hypersetup{citecolor={blue}}
\usepackage{txfonts}
\usepackage{float} 

\begin{document}

   \title{Metallicity-dependent kinematics and orbits in the Milky Way's nuclear stellar disc}

  \author{F. Nogueras-Lara
          \inst{1}       
           \and      
           N. Nieuwmunster       
          \inst{2,3} 
          \and
          M. Schultheis
          \inst{2}
            \and
          M. C. Sormani
          \inst{4}
            \and
          F. Fragkoudi
          \inst{5}
          \and
          B. Thorsbro
          \inst{2,3}
          \and
          R.\,M.\,Rich
          \inst{6}
          \and
          N. Ryde
          \inst{3}
          \and
          J. L. Sanders
          \inst{7}          
          \and
          L. C. Smith   
         \inst{8}           
          }

   \institute{
    European Southern Observatory, Karl-Schwarzschild-Strasse 2, D-85748 Garching bei M\"unchen, Germany
              \email{francisco.nogueraslara@eso.org}                         
     \and   
       Université Côte d'Azur, Observatoire de la Côte d'Azur, Laboratoire Lagrange, CNRS, Blvd de l'Observatoire, F-06304 Nice, France
     \and
       Division of Astrophysics, Department of Physics, Lund University, Box 43, SE-22100 Lund, Sweden
      \and
      Department of Physics, University of Surrey, Guildford GU2 7XH, UK
      \and
      Institute for Computational Cosmology, Department of Physics, Durham University, South Road, Durham DH1 3LE, UK
      \and
      Department of Physics and Astronomy, UCLA, 430 Portola Plaza, Box 951547, Los Angeles, CA 90095-1547, USA
      \and 
      Department of Physics and Astronomy, University College London, London WC1E 6BT, UK
      \and
      Institute of Astronomy, University of Cambridge, Madingley Rd, Cambridge CB3 0HA, UK      
      }

   \date{}

 
  \abstract
   {The nuclear stellar disc (NSD) is a flat and dense stellar structure at the centre of the Milky Way. Previous work has identified the presence of metal-rich and metal-poor stars in the NSD, suggesting that they have different origins. The recent publication of photometric, metallicity, proper motion, and orbital catalogues allows the NSD stellar population to be characterised with unprecedented detail.}
   {We aim to explore the proper motions and orbits of NSD stars with different metallicities to assess whether they have different origins and to better understand the metallicity distribution in the NSD.}
   {We distinguished between metal-rich and metal-poor stars by applying a Gaussian mixture model, as done in previous work, and analysed the proper motions, orbits, and spatial distribution of stars with different metallicities.}
   {We find that metal-rich stars exhibit a lower velocity dispersion, suggesting that they trace a kinematically cooler component compared to metal-poor ones. Furthermore, z-tube orbits are predominant among metal-rich stars, while chaotic/box orbits are more common among metal-poor ones. We also find that metal-rich and metal-poor stars show a similar extinction and are present throughout the analysed regions. As a secondary result, we detected a metallicity gradient in the metal-rich population with higher metallicity towards the centre of the NSD and a tentative gradient for the metal-poor stars, which is consistent with previous studies that did not distinguish between the two populations.} 
   {Our results suggest that metal-rich stars trace the NSD, whereas metal-poor ones are related to the Galactic bar and probably constitute Galactic bar interlopers and/or are NSD stars that originated from accreted clusters. The detected metallicity gradients aligns with the currently accepted inside-out formation of the NSD.}

   \keywords{Galaxy: nucleus -- Galaxy: centre -- Galaxy: structure  -- Galaxy: bulge -- Galaxy: stellar content -- Galaxy: kinematics and dynamic -- proper motions -- dust, extinction
               }

   \maketitle%

\section{Introduction}

The heart of the Milky Way stands out as our closest galaxy centre, and it is the only one where we can study stars down to milli-parsec scales \citep[e.g.][]{Dong:2011ff,Nogueras-Lara:2019aa,Do:2019aa,Gravity-Collaboration:2020aa}. Therefore, it constitutes a unique template for analysing the stellar population of a galaxy nucleus, which can provide a better understanding of galactic nuclei and their role in galaxy formation and evolution. Nevertheless, high extinction and extreme source crowding hamper analysis of the Galactic centre stellar population and primarily limit it to the near-infrared regime \citep[e.g.][]{Nishiyama:2006tx,Nishiyama:2008qa,Schultheis:2009tg,Fritz:2011fk,Alonso-Garcia:2017aa,Thorsbro:2020uq,Nogueras-Lara:2021wj,Sanders:2022wa}. 

\begin{figure*}
                 
   \includegraphics[width=\linewidth]{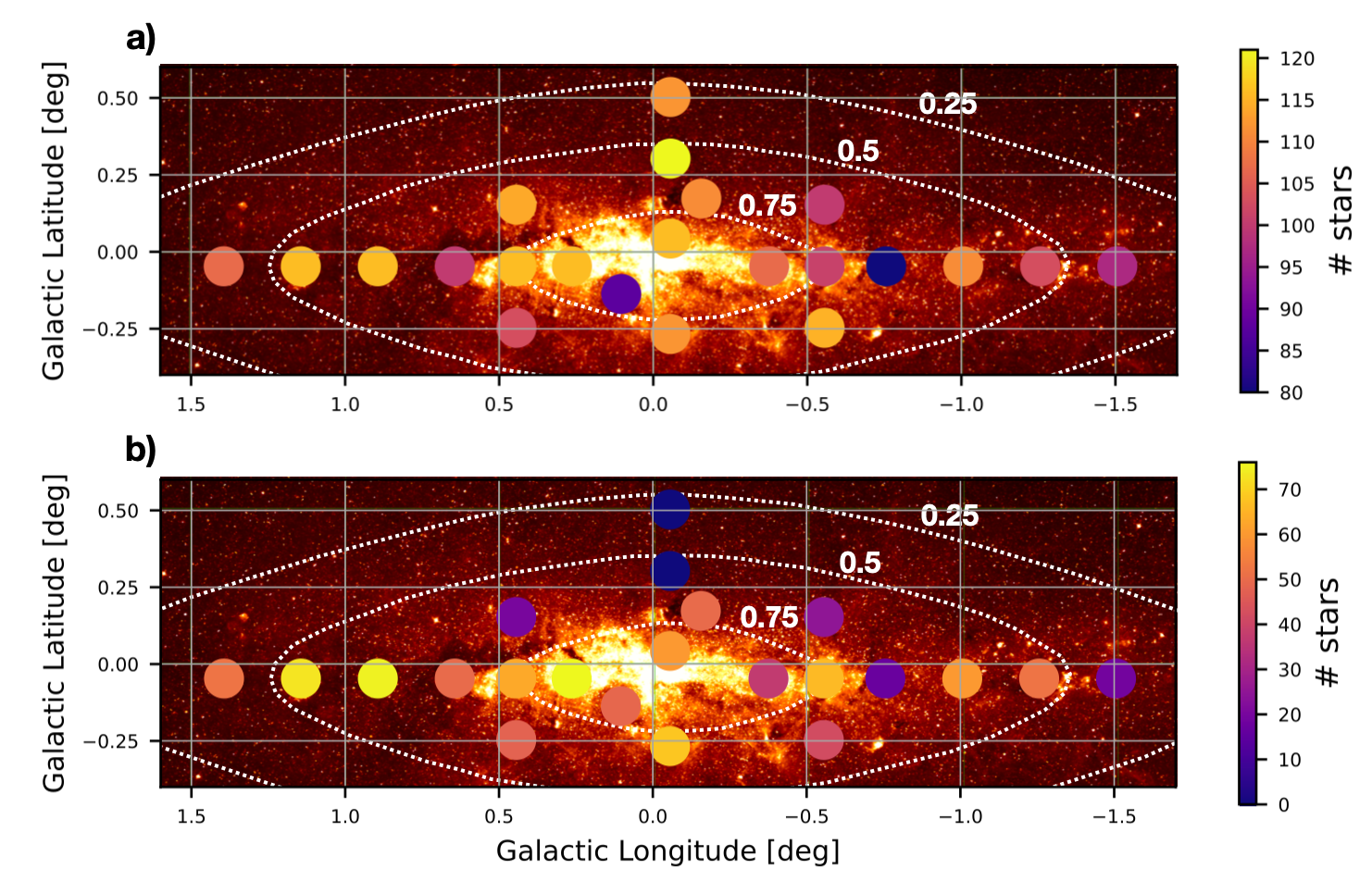}
   \caption{Scheme of the observed region overplotted on a Spitzer/IRAC image at 8\,$\mu$m. \citep{Stolovy:2006fk}. This wavelength highlights the bright and dusty clouds that outline the NSD. a) KMOS fields in which the colour scale indicates the star counts. b) Number of stars per field with computed orbits \citep{Nieuwmunster:2024aa}. The two northernmost fields do not contain any stars due to the selection criteria used to create the orbital catalogue \citep[$|l|$ < $1.5^\circ$ and |b| < $0.3^\circ$; for further details, see][]{Nieuwmunster:2024aa}. The dotted contours show the 25\%, 50\%, and 75\% levels of the NSD self-consistent model obtained by \citet{Sormani:2022wv}.}

\label{scheme}
\end{figure*}

The Galactic centre is roughly outlined by the nuclear stellar disc (NSD), a flat and likely axisymmetric stellar structure with a scale length of $\sim100$\,pc and a scale height of $\sim40$\,pc \citep[e.g.][]{gallego-cano2019,Nogueras-Lara:2022aa,Sormani:2022wv}. It contains a total stellar mass of $\sim 10^9$\,M$_\odot$ \citep{Launhardt:2002nx,Sormani:2022wv} and is characterised by a predominantly old stellar population and a bursty star-formation history \citep[e.g.][]{Nogueras-Lara:2019ad,Sanders:2022ab,Schodel:2023aa}, which makes it different from the surrounding and much larger Galactic bar/bulge \citep[e.g.][]{Zoccali:2003aa,Valenti:2013aa,Wegg:2013kx,Wegg:2015aa,Portail:2017aa,Surot:2019ab}. Additionally, the NSD contains  $\sim10^6$\,M$_\odot$ of young stars that formed in situ in the last 30\,Myr \citep[e.g.][]{Matsunaga:2013uq,Nogueras-Lara:2019ad,Nogueras-Lara:2022ua,Nogueras-Lara:2024aa}, and it has contributed to up to 10\% of the star-forming activity observed throughout the entire Milky Way over the past $\sim100$\,Myr \citep[e.g.][]{Mezger:1996uq,Mauerhan:2010jc,Matsunaga:2011uq,Crocker:2011kx,Nogueras-Lara:2019ad}.

The NSD is a rotating structure \citep{Lindqvist:1992fk,2015ApJ...812L..21S,Fritz:2020aa,Shahzamanian:2021wu} that appears to be kinematically distinct from the Galactic bar/bulge \citep{2015ApJ...812L..21S,Schultheis:2021wf}. However, \citet{Zoccali:2024aa} recently reported that they were unable to definitively identify the NSD as a distinct kinematic feature when comparing it to nearby control fields in the Galactic bar/bulge, contradicting previous work. Their analysis used proper motions derived from the VISTA Variables in the Vía Láctea survey \citep{Minniti:2010fk} and APOGEE DR17 data \citep{Majewski:2017aa,Abdurrouf:2022aa}.

The NSD stellar population exhibits a broad metallicity distribution that has been previously fitted using two Gaussian models accounting for metal-rich and metal-poor stars, respectively \citep{Schultheis:2021wf,Nogueras-Lara:2022tp,Nogueras-Lara:2023ab}. The metal-rich component (mean metallicity $[M/H] = 0.12$, \citealt{Schultheis:2021wf}) constitutes approximately 70-80\% of the NSD's stars and shows a lower line-of-sight velocity dispersion in comparison with the metal-poor component (mean metallicity $[M/H] = -0.22$, \citealt{Schultheis:2021wf}). This suggests that metal-rich stars have a different origin and probably formed in situ \citep{Schultheis:2021wf} from the dense accumulation of gas and dust that comprises the central molecular zone \citep[CMZ; e.g.][]{Henshaw:2022vl}. This is further supported by indications that stars in the NSD and the gas in the CMZ appear to be co-rotating \citep{2015ApJ...812L..21S,Schultheis:2021wf}.


In this paper, we aim to study the proper motion distribution of NSD stars with different metallicities to assess previous analysis using line-of-sight velocities and to confirm whether they have different kinematics and origin. Additionally, we utilise the recent orbital analysis of NSD stars by \citet{Nieuwmunster:2024aa} to determine whether stars with different metallicities have different orbits. Analysing stars with different kinematics and orbits also helps shed light on the potential kinematic difference between the NSD and the Galactic bar/bulge, and it is crucial to understanding their relation as well as the NSD formation mechanism.



\section{Data}


\subsection{Metallicity, photometry, and proper motions}
\label{target}

We used the spectroscopic survey of the NSD \citep{Fritz:2020aa} obtained with the KMOS instrument at the VLT \citep{Sharples:2013aa}. The chosen sample includes the primary targets of the survey contained within the central region of the NSD shown in Fig.\,\ref{scheme} (2303 stars). The metallicities were derived with an uncertainty of $\sim0.3$\,dex and a calibration uncertainty of $\sim 0.1$\,dex \citep{Fritz:2020aa}.

We also used $H$ and $K_s$ photometry from the SIRIUS/IRSF survey of the Galactic centre \citep{Nagayama:2003fk,Nishiyama:2006tx,Nishiyama:2013uq}, which is available for all the targets of the KMOS spectroscopic survey of the NSD \citep[see Table\,E1 in][]{Fritz:2020aa}. The mean zero-point uncertainties are $\sim0.03$\,mag in both bands, and the 10\,$\sigma$ limiting magnitudes are $H=16.6$\,mag and $K_s=15.6$\,mag.

For the kinematical analysis, we used proper motions from a preliminary version of the second version of VIRAC  \citep[VIRAC2,][]{Smith_submitted}. This data set was also used by \citet{Sormani:2022wv} and \citet{Nieuwmunster:2024aa}. It contains absolute proper motions of stars in the spectroscopic survey, using Gaia as an astrometric reference. \citet{Sormani:2022wv} describes the data set in more detail.

We reduced the contamination from foreground stars (mainly from the spiral arms along the line of sight towards the Galactic centre and up to a certain extent to the Galactic bar/bulge) by using their significantly different extinction \citep[e.g.][]{Nogueras-Lara:2021uz}. We excluded them in all fields by applying a colour cut, ($H-K_s$) > max(1.3, -0.0233 $K_s$+1.63), as done in \cite{Schultheis:2021wf} and \citet{Nieuwmunster:2024aa}. Additionally, we removed stars with high proper motion uncertainties ($>1$\,mas/yr). This can lead to an effective magnitude cut at $K\sim10$\,mag (Fig.\,\ref{CMD}) because the uncertainties are high for too bright stars \citep[see Sect.\,3 in ][]{Sormani:2022wv}. We ended up with a total of 1497 stars in the target region for our subsequent analysis.

\subsection{Orbital catalogue}

We utilised the orbital catalogue obtained by \citet{Nieuwmunster:2024aa}, which was derived from the previously described KMOS spectroscopic survey of the NSD \citep{Fritz:2020aa} and the preliminary proper motions from VIRAC2 \citep{Smith_submitted}. The orbital catalogue also implemented the previously mentioned colour cut to mitigate potential contamination from foreground sources and applied a strict proper motion uncertainty cut, $d\mu_{l,b}<0.6$\,mas/yr, resulting in a sample of 1130 stars. \citet{Nieuwmunster:2024aa} used $K_s$ luminosity functions to estimate the completeness of the data to be $\sim15$\% at $K_s\sim15$\,mag. Thus, the stellar sample is limited, and the fraction of star orbits may not fully represent the overall NSD stellar population.

The orbits of the stars were integrated in a non-axisymmetric potential, combining the potentials corresponding to the Galactic bar/bulge density \citep{Launhardt:2002nx}, the NSD \citep{Sormani:2022wv}, and the nuclear star cluster \citep{Chatzopoulos:2015yu}. They also assumed a distance distribution considering a mean Galactic centre distance of $8.2$\,kpc and an NSD scale length of $\sim100$\,pc \citep[e.g.][]{gallego-cano2019,Sormani:2022wv,Nogueras-Lara:2022aa}. The final catalogue comprises several orbital families (i.e. chaotic, z-tube, x-tube, banana, fish, saucer, pretzel, 5:4, and 5:6 orbits), which are present in the NSD region. Examples of each of the orbital families are shown in Appendix\,A of \citet{Nieuwmunster:2024aa}.

Figure\,\ref{CMD} shows a colour magnitude diagram of $K_s$ versus $H-K_s$ of the observed fields. The parallelogram-shaped area is due to the selection criteria for the target selection of the KMOS survey. The targets were chosen to have an extinction-free apparent magnitude in the range $K_{0}\in[7.0,9.5]$\,mag \citep[for further details, see][]{Fritz:2020aa}.

\begin{figure}
                 
   \includegraphics[width=\linewidth]{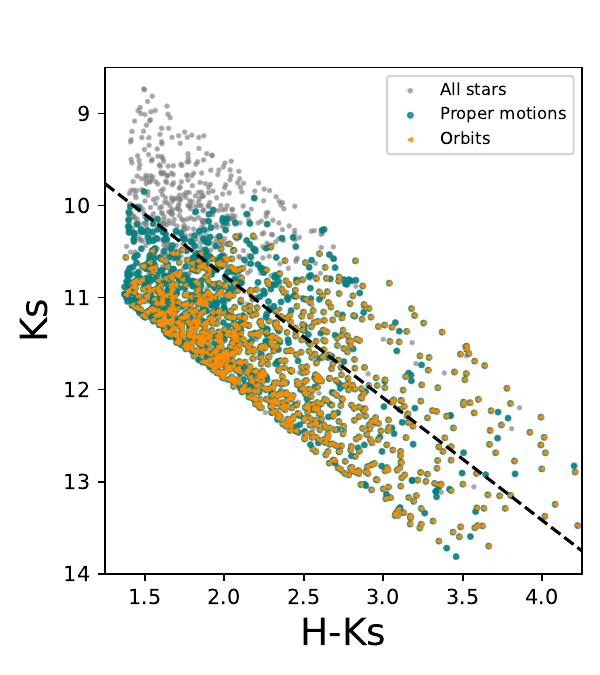}
   \caption{Colour-magnitude diagram $K_s$ versus $H-K_s$. The grey dots represent all the stars in the KMOS catalogue in the analysed region after excluding the foreground population. Stars with proper motions below 1\,mas/yr and stars with orbits are overplotted with different colours and symbols, as indicated in the legend. The black dashed line corresponds to the cut applied in Sects.\,\ref{bimodal}, \ref{met_tube} and \ref{extinction} to analyse the effect of missing bright stars at $K_s\sim10$\,mag when applying the proper motion quality cut.}

\label{CMD}
\end{figure}

\section{Metallicity distribution and proper motions}

\label{bimodal}
\begin{figure}
                 
   \includegraphics[width=\linewidth]{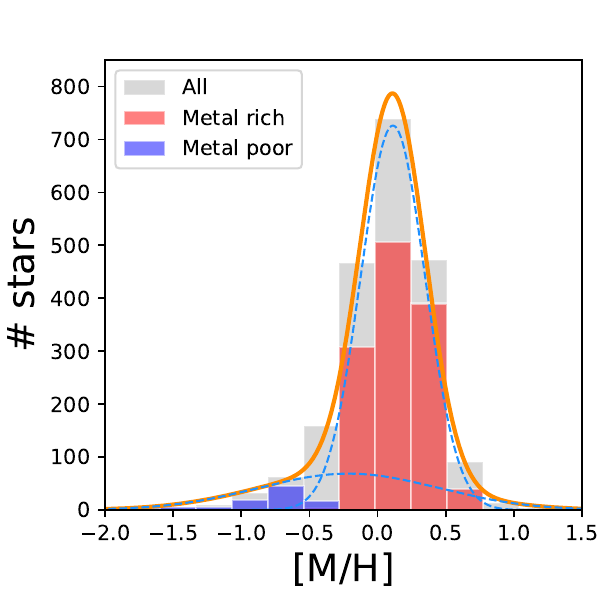}
   \caption{Two-Gaussian decomposition of the NSD metallicity distribution. The orange line shows the result of the GMM analysis, whereas the blue dashed lines depict each of the Gaussian components. Red and blue bars correspond to stars with a more than 70\% probability of belonging to each of the components.}

\label{met_all}
\end{figure}

To investigate the proper motion distribution of stars with different metallicities, we first analysed the metallicity distribution of the target stars by repeating the analysis in  \citet{Schultheis:2021wf}. Namely, we applied a Gaussian mixture model \citep[GMM;][]{Pedregosa:2011aa} to obtain the number of Gaussians that best reproduce the metallicity distribution. We distinguished between single-Gaussian and two-Gaussian  models by applying the Bayesian information criterion \citep{Schwarz:1978aa} and the Akaike information criterion \citep{Akaike:1974aa}. We found that the data is best represented by a two-Gaussian model. Figure\,\ref{met_all} shows the obtained results. To estimate the mean metallicity and the associated uncertainty for each component, we resorted to Monte Carlo (MC) simulations and generated 1000 MC samples by randomly varying the metallicity of each star, assuming Gaussian uncertainties. We performed the GMM analysis on each of the 1000 samples, assuming a two-Gaussian distribution, and computed the final values along with their associated uncertainties as the mean and standard deviation. The results, shown in Table\,\ref{met_all_table} are in perfect agreement with those reported by \citet{Schultheis:2021wf}.

\setlength{\tabcolsep}{3.5pt}
\def\arraystretch{1.1}

\begin{table*}

\caption{Characterisation of the two-Gaussian decomposition of the NSD metallicity distribution.}
\label{met_all_table} 
\begin{center}
  
\begin{tabular}{cccc|cccc|cc}
\hline 
\hline 
 & $W$ & $[M/H]$ & $\sigma _{[M/H]}$ & $\mu_l$ & $\sigma_{\mu_l}$ & $\mu_b$ & $\sigma_{\mu_b}$ & $H-K_s$& $\sigma_{H-K_s}$\tabularnewline
\hline 
MR & 0.79 $\pm$ 0.02 & 0.11 $\pm$ 0.01 & 0.26 $\pm$ 0.01 & ${-1.13}$ $\pm$ 0.08 & 2.70 $\pm$ 0.05 & 0.17 $\pm$ 0.05 & ${1.88}$ $\pm$ 0.05 & 2.24 $\pm$ 0.02 & 0.57 $\pm$ 0.01\tabularnewline
MP & 0.21 $\pm$ 0.02 & ${-0.19}$ $\pm$ ${0.03}$ & ${0.73}$ $\pm$ 0.04 & ${-1.43}$ $\pm$ ${0.31}$ & ${3.17}$ $\pm$ 0.21 & ${-0.01}$ $\pm$ 0.32 & ${3.34}$ $\pm$ ${0.27}$ & 2.21 $\pm$ ${0.06}$ & 0.59 $\pm$ 0.04\tabularnewline
\hline 
 &  &  & \multicolumn{1}{c}{} &  &  &  & \multicolumn{1}{c}{}{} & \tabularnewline
\end{tabular}
    
\end{center}
\footnotesize
\textbf{Notes.} $W$, $[M/H]$, and $\sigma_{[M/H]}$ correspond to the results from the GMM analysis of the two-Gaussian distribution, where $W$ indicates the relative weight of each of the Gaussian components. We note that $\mu_l$, $\sigma_{\mu_l}$, $\mu_b$, and $\sigma_{\mu_b}$ are in mas/yr, and $H-K_s$ and $\sigma_{H-K_s}$ are in Vega magnitudes.

\end{table*}

We also employed the GMM approach to calculate the likelihood of each star being associated with each of the Gaussian distributions. To analyse the proper motion distribution of stars belonging to each metallicity component, we assumed that a star belongs to either the metal-rich or the metal-poor component if its likelihood of being associated with that component is larger than 70\%. Given that the stellar metallicity of the KMOS catalogue was calibrated using empirical spectra with $[M/H] < 0.6$\,dex, we excluded stars with higher metallicities from the subsequent analysis. Figure\,\ref{met_all} shows the final selection of stars. We ended up with 1249 metal-rich and 106 metal-poor stars.

\begin{figure}
                 
   \includegraphics[width=\linewidth]{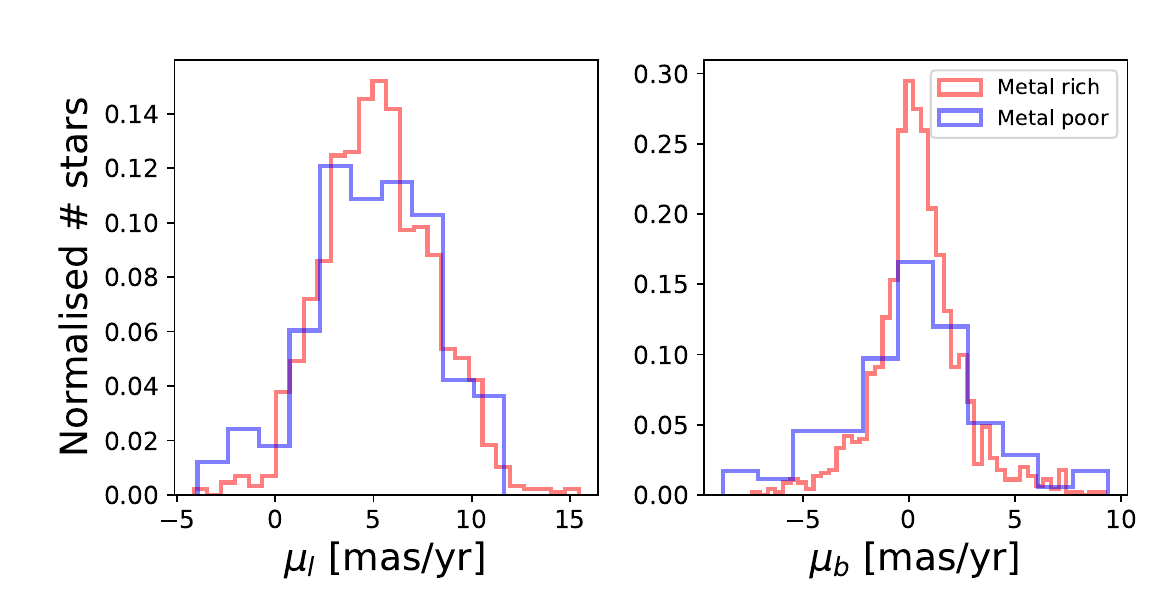}
   \caption{Proper motion distribution of metal-rich and metal-poor stars in the NSD following the GMM analysis (see Sect.\,\ref{bimodal}).}

\label{proper_all}
\end{figure}

We built histograms for the proper motion component parallel ($\mu_l$) and perpendicular ($\mu_b$) to the Galactic plane. To estimate the mean and the standard deviation of each distribution and their corresponding uncertainties, we applied a 3$\sigma$ clipping algorithm to remove potential outliers, and we used a bootstrap resampling method with 5000 iterations. Table\,\ref{met_all_table} and Fig.\,\ref{proper_all} show the obtained results. To assess potential sources of systematic uncertainties, we repeated the calculation, considering stars with uncertainties below $1.2$\,mas/yr and $0.8$\,mas/yr as well as by adjusting the cut to remove the foreground population (i.e. $H-K_s\sim1.1$\,mag).  We also repeated the calculation using all stars within the colour cut $H-K_s\gtrsim1.3$\,mag and with proper motion uncertainties below 1\,mas/yr, assuming that stars with $[M/H]<-0.25$\,dex are metal poor, while those with $-0.25$\,dex$<[M/H]<0.6$\,dex are metal-rich. Finally, we explored the effect of the effective cut at $K_s\sim10$\,mag when considering stars with proper motion uncertainties below 1\,mas/yr (see Fig.\,\ref{CMD}). For this, we defined a new selection cut parallel to the original one \citep[following the reddening vector, ][]{Fritz:2011fk} in order to exclude the stars with $K_s>10$\,mag, keeping only the ones below the black dashed line in Fig.\,\ref{CMD}. In all cases, the results were consistent within the uncertainties.

We concluded that although the mean proper motions are similar for both stellar populations, there is a significant difference in the velocity dispersion of the $\mu_b$ component. Namely, the metal-poor component exhibits a $\mu_b$ standard deviation that is approximately a factor of two larger than that of the metal-rich group. Regarding the $\mu_l$ component, we found that the velocity dispersion is slightly larger for the metal-poor stellar population, although there is not a clearly significant difference. This is not surprising because the proper motion component parallel to the Galactic plane is much more affected by the extinction along the line of sight and because of the fact that the NSD is a rotating structure. Therefore, the presence of stars moving eastwards and westwards depends on the completeness of the data and will be different for the different regions analysed (e.g. \citealt{2015ApJ...812L..21S,Martinez-Arranz:2022uf,Nogueras-Lara:2022aa} and Figs.\,A1-A4 in \citealt{Sormani:2022wv}, which show the $\mu_l$ and $\mu_b$ decomposition between the Galactic bar and the NSD). 

\begin{figure}
                 
   \includegraphics[width=\linewidth]{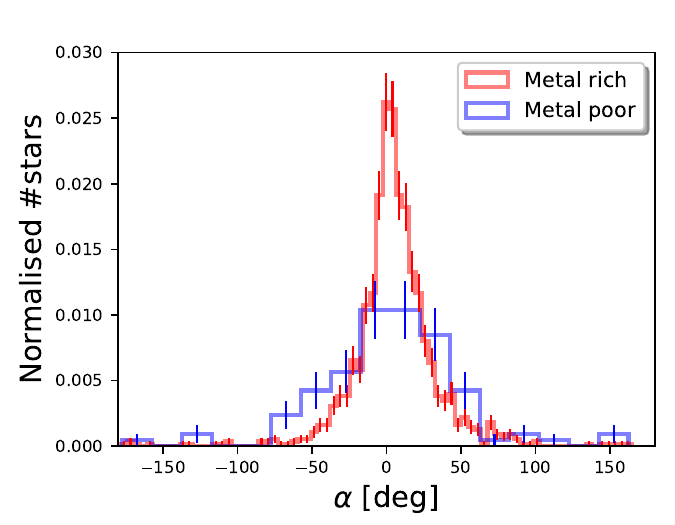}
   \caption{Proper motion orientation of metal-rich and metal-poor stars. The uncertainties correspond to Poisson errors (i.e. $\sqrt{\text{\#stars/bin}}$) scaled to the normalised number of stars in each bin.}

\label{inclination}
\end{figure}

We also explored the proper motion orientation, $\alpha = \text{atan}(\mu_b/\mu_l)$, of metal-rich and metal-poor stars. We subtracted the proper motion velocity of Sgr\,A* \citep[$\mu_l = -6.396\pm0.071$\,mas/yr and $\mu_b = -0.239\pm0.045$\,mas/yr, see ][]{Gordon:2023aa} to obtain Galactic proper motions relative to the Galactic centre. To ensure $\alpha \in [-180,180]$\,degrees, we adjusted the output based on the orientation of the proper motions and adopted the origin pointing east. Figure\,\ref{inclination} shows the obtained results. We computed the mean orientation values, their standard deviations, and the associated uncertainties by applying a bootstrap resampling method with 5000 iterations. We obtained $\alpha_{rich} = 5\pm1^\circ$, $\sigma_{\alpha_{rich}} = 30\pm1^\circ$, and $\alpha_{poor} = {3\pm5}^\circ$, $\sigma_{\alpha_{poor}} = {48}\pm 5 ^\circ$, for the metal-rich and the metal-poor component, respectively. Although the orientation values for both populations are compatible, the metal-poor stellar population shows a significantly larger standard deviation.

\section{Metallicity distribution and orbits}

\subsection{Orbits for each metallicity group}

We used the orbital catalogue obtained by \citet{Nieuwmunster:2024aa} to explore whether stars with different metallicities and kinematics undergo different orbits. The authors used a statistical approach to obtain the orbits, and it may not be valid on an individual star basis. However, this does not pose any problem because our objective is to understand the mean behaviour of stars with different metallicities. \citet{Nieuwmunster:2024aa} found that z-tube orbits (including all orbital families with a circulation around the z-axis, such as z-tube, fish, saucer, pretzel, 5:4, and 5:6 orbits) dominate the orbital families, comprising approximately 70\% of the stars in the catalogue, and they are directly associated with the NSD. Conversely, chaotic/box orbits constitute about 25\% of the sample and are likely related to Galactic bar orbits.

\begin{figure}
                 
   \includegraphics[width=\linewidth]{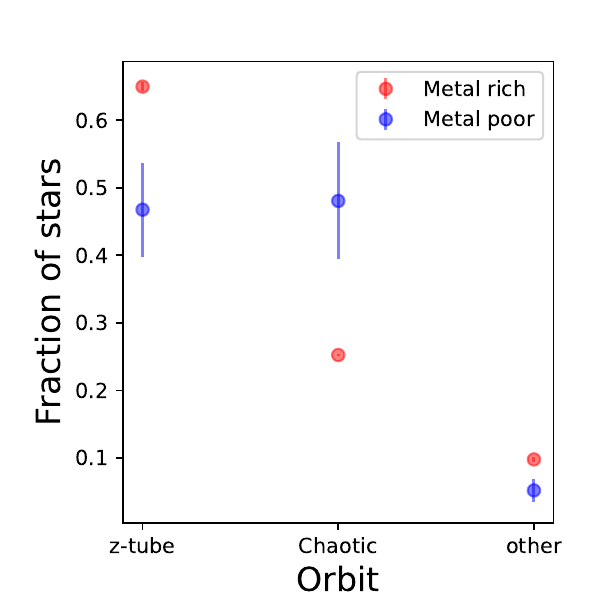}
   \caption{Relative fraction of stars with z-tube and chaotic/box orbits for the metal-rich and metal-poor stellar populations defined in Sect.\,\ref{bimodal}. The label "other" corresponds to stars following banana and x-tube orbits (see \citet{Nieuwmunster:2024aa} for further details). The uncertainty bars for the metal-rich population are smaller to the red circles and are therefore not visible in the figure.}

\label{orbits_all}
\end{figure}

We analysed the fraction of stars with these orbits belonging to the previously defined metal-rich and metal-poor stellar populations. We obtained that the orbits are different for the different metallicity components, as shown in Fig.\,\ref{orbits_all}. The uncertainties were estimated based on an assumed membership probability of 60\% and 80\% for each of the metallicity groups, while the mean values correspond to the 70\% probability. We concluded that z-tube orbits dominate the metal-rich component ($\sim70$\% of the metal-rich stars belong to this orbital family) and are less frequent in the metal-poor stellar population, in which the chaotic/box orbits dominate and correspond to $\sim50$\% of the stars with defined orbits in the metal-poor component.

\subsection{Metallicity for z-tube and chaotic/box orbits}
\label{met_tube}

\begin{figure}
                 
   \includegraphics[width=\linewidth]{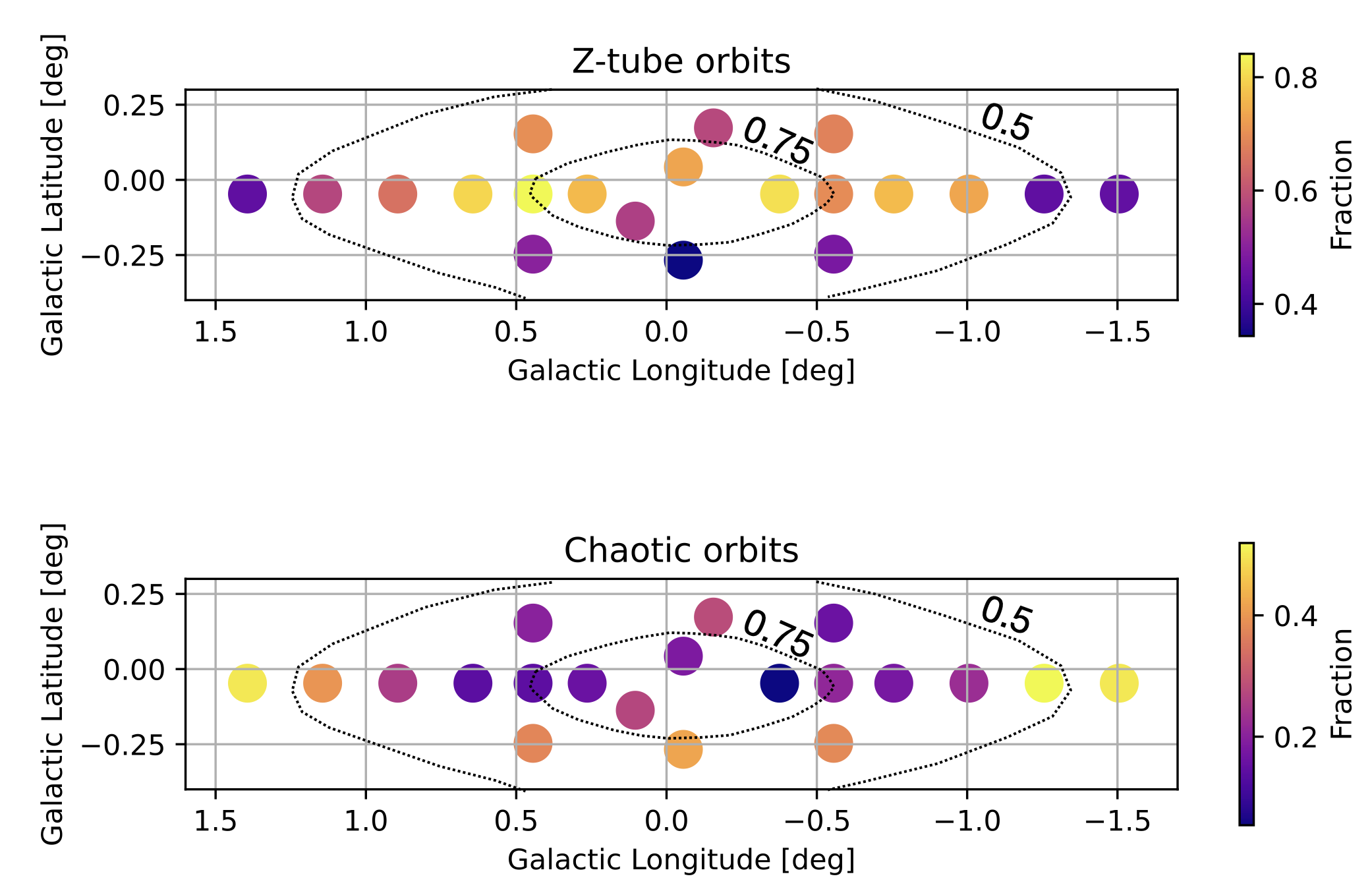}
   \caption{Relative fraction of stars with z-tube and chaotic orbits. The contours show the 50\% and 75\% levels of the NSD self-consistent model
obtained by \citet{Sormani:2022wv}.}

\label{orbits}
\end{figure}

We also explored the metallicity distribution of all the stars in the catalogue of orbits with z-tube and chaotic/box orbits. To begin, we analysed the spatial distribution of z-tube and chaotic/box orbits. Figure\,\ref{orbits} shows the obtained results. We found that the relative fraction of stars undergoing chaotic orbits is more common towards the edge of the NSD, as was also discussed by \citet{Nieuwmunster:2024aa}, supporting their relation with potential contaminants from the bar. In contrast, stars with z-tube orbits and, in general, stars without chaotic orbits, are more centrally concentrated and seem to better trace the NSD.

\begin{figure}
                 
   \includegraphics[width=\linewidth]{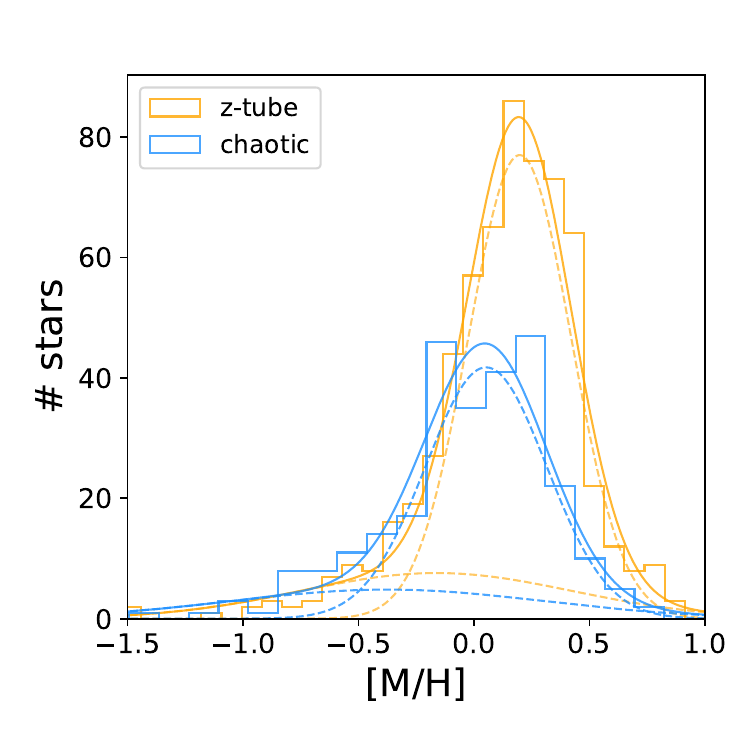}
   \caption{Metallicity distribution of all stars in the orbital catalogue \citep{Nieuwmunster:2024aa} following z-tube orbits (including all orbital families with a circulation around the z-axis) and chaotic/box orbits. The solid lines show the result of the GMM analysis, whereas the dashed lines correspond to each Gaussian model.}

\label{bimodal_orbits}
\end{figure}

We computed the metallicity distribution for stars with z-tube and chaotic/box orbits and observed that they are different, with z-tube stars on average being more metal rich (Fig.\,\ref{bimodal_orbits}). We also applied the previously described GMM approach (see Sect.\,\ref{bimodal}) to compare a single- and a two-Gaussian model. In all cases, we found that a two-Gaussian model is preferred. Figure\,\ref{bimodal_orbits} shows the results for each of the orbital families. We also computed the parameters associated with the metal-rich and metal-poor stellar populations by resorting to MC simulations with 1000 iterations, as explained in Sect.\,\ref{bimodal}. The results are presented in Table\,\ref{met_orbits_table}. We examined the effect of potential systematic uncertainties related to the effective cut at $K_s\sim10$\,mag resulting from the exclusion of stars with proper motion uncertainties above 0.6\,mas/yr in the orbital catalogue \citep{Nieuwmunster:2024aa}. For this, we repeated the analysis, applying the cut shown in Fig.\,\ref{CMD} as previously explained in Sect.\,\ref{bimodal}. We did not find any significant difference within the uncertainties.

\begin{table}

\caption{Characterisation of the two-Gaussian metallicity decomposition of stars with different orbits.}
\label{met_orbits_table} 
\begin{center}
\def\arraystretch{1.3}
\setlength{\tabcolsep}{1.9pt}

\begin{tabular}{ccccc}
 &  &  &  & \tabularnewline
\hline 
\hline 
Orbit & Metallicity & $W$ & $[M/H]$ & $\sigma_{[M/H]}$\tabularnewline
\hline 
z-tube & Rich & ${0.80}$ $\pm$ 0.05 & 0.19 $\pm$ 0.01 & 0.26 $\pm$ 0.01\tabularnewline
 & Poor & ${0.20}$ $\pm$ 0.05 & ${-0.16}$ $\pm$ ${0.16}$ & 0.67 $\pm$ 0.09\tabularnewline
\hline 
Chaotic & Rich & 0.78 $\pm$ ${0.06}$ & 0.05 $\pm$ 0.02 & ${0.29}$ $\pm$ 0.03\tabularnewline
 & Poor & 0.22 $\pm$ ${0.06}$ & ${-0.49}$ $\pm$ ${0.43}$ & 0.73 $\pm$ ${0.12}$\tabularnewline
\hline 
 &  &  &  & \tabularnewline
\end{tabular}

\end{center}
\footnotesize
\textbf{Notes.} $W$, $[M/H]$, and $\sigma_{[M/H]}$ correspond to the results from the GMM analysis of the two-Gaussian distribution, where $W$ indicates the relative weight of each of the Gaussian components.

\end{table}

We observed that stars exhibiting chaotic/box orbits, regardless of their metallicity, are generally more metal poor compared to those following z-tube orbits (Fig.\,\ref{bimodal_orbits}). The mean metallicity values obtained for both metallicity components with chaotic/box orbits align well with the results from \citet{Schultheis:2021wf} for the inner regions of the Galactic bar/bulge (0.04 and -0.55\,dex for the metal-rich and metal-poor components, respectively) based on analysis of APOGEE data \citep{Ahumada:2020aa,Rojas-Arriagada:2020aa}. Similarly, the metallicity values obtained for stars undergoing z-tube orbits closely match our results for the NSD (see Table\,\ref{met_all_table}).

\section{Extinction of stars with different metallicities along the line of sight}
\label{extinction}

The NSD formed $\gtrsim8$\,Gyr ago and appears to be dominated by an old stellar population \citep[$>80$\% of the stellar mass is older than $\gtrsim7$\,Gyr, e.g.][]{Nogueras-Lara:2019ad,Nogueras-Lara:2022ua,Sanders:2022ab,Schodel:2023aa}. Assuming that metal-rich and metal-poor stars have a similar age allowed us to use the $H-K_s$ colour as a proxy for the extinction towards these stars \citep[e.g.][]{Nogueras-Lara:2018aa,Nogueras-Lara:2021wj}. Moreover, there is a statistical correlation between distance along the line of sight and extinction in the NSD \citep[e.g.][]{Nogueras-Lara:2022aa,Nogueras-Lara:2023aa}, which makes it possible to estimate whether the spatial distribution is similar for both stellar populations by analysing their extinction. We would therefore expect a significant difference in extinction if one of the components is located in a specific region along the line of sight and thus the stars would not be homogeneously distributed.

\begin{figure}
                 
   \includegraphics[width=\linewidth]{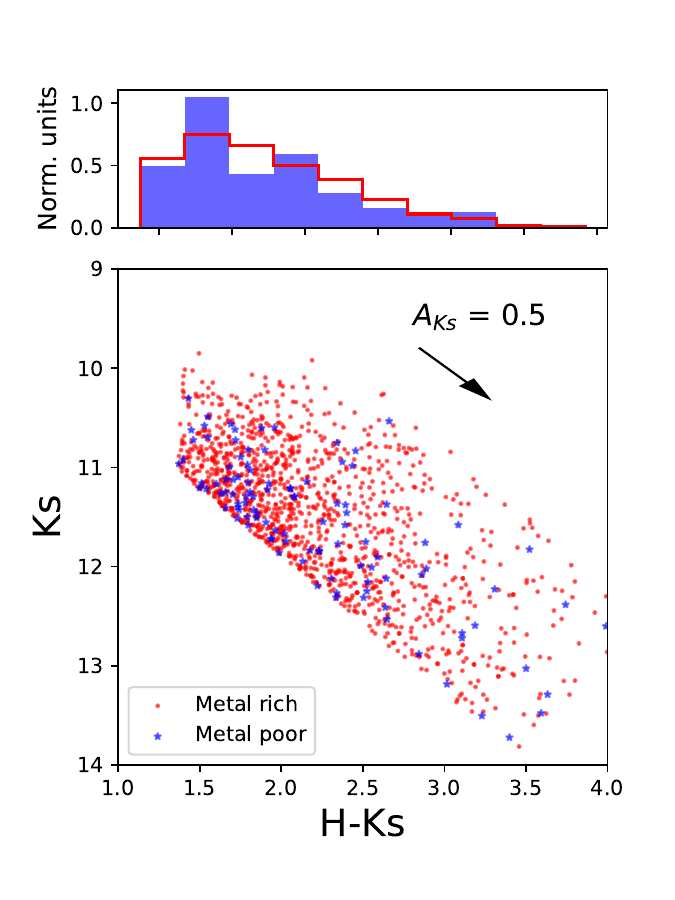}
   \caption{Colour-magnitude diagram of the metal-rich and metal-poor stars used in Sect.\,\ref{bimodal}. The black arrow shows the direction of the reddening vector.}

\label{ext}
\end{figure}

We built a colour-magnitude diagram $K_s$ versus $H-K_s$ to analyse the colour distribution (Fig.\,\ref{ext}) of the stars belonging to each stellar population whose proper motions were previously analysed. We computed the mean colour and its associated uncertainty for the metal-rich and the metal-poor stars by applying a bootstrap resampling method with 5000 iterations. Table\,\ref{met_all_table} shows the obtained results. We found that the metal-poor stars show a somewhat smaller $H-K_s$ colour, which might indicate that they are more abundant close to the NSD edge. Nevertheless, the colours for both stellar populations (and their standard deviations) are similar within the uncertainties, so there is no clear difference in the spatial distribution along the line of sight. We detected metal-poor and metal-rich stars with both large and small extinction, as indicated by the $H-K_s$ standard deviations, showing that both stellar populations are present all over the NSD. We assessed potential systematic uncertainties as previously explained in Sect.\,\ref{bimodal}. Namely, we varied the proper motion uncertainty cut to 0.8\,mas/yr and 1.2\,mas/yr, we modified the foreground cut to $H-K_s\sim1.1$\,mag, and we applied a more restrictive selection cut, as indicated in Fig.\,\ref{CMD}. The results were always consistent within the uncertainties, with the only difference being a somewhat smaller extinction for both stellar populations when assuming the more restrictive selection criterion. In any case, this was expected because the new cut removes the most reddened stars.

\section{Galactic longitude analysis of stars with different metallicities}
\label{spat}

We studied the metallicity distribution of KMOS stars at different longitudes from the Galactic centre. In this way, we investigated whether the two-Gaussian metallicity profile is homogeneously distributed across the field and whether metal-rich and metal-poor stars have different spatial distributions.

\begin{figure}
                 
   \includegraphics[width=\linewidth]{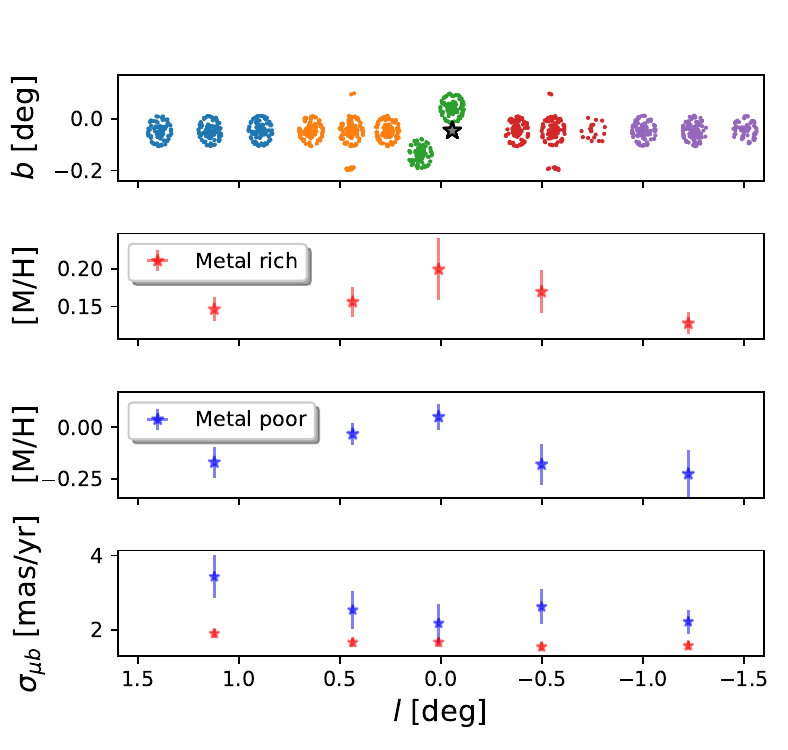}
   \caption{Line-of-sight analysis of the NSD metallicity and kinematics (the corresponding values are included in Table.\,\ref{spatial_gaussian}). Top panel: Target KMOS stars divided into five colour-coded regions at different longitudes. The grey star indicates the position of Sgr\,A*. Middle panels: Mean metallicity obtained for the metal-rich (top) and the metal-poor (bottom) components of the two-Gaussian distribution. We plot metal-rich and metal-poor stars separately to highlight the detected gradients more effectively. Bottom panel: Velocity dispersion of $\mu_b$ for the metal-rich and metal-poor components.}

\label{spatial}
\end{figure}

To avoid mixing stellar populations at significantly different latitudes, whose metallicity might also vary vertically, we restricted our analysis to stars within a latitude of $\pm20$\,pc from the centre of the NSD (Sgr\,A*), as shown in Fig.\,\ref{spatial}. For our analysis, we defined five regions with different longitudes, as colour-coded in the upper panel of Fig.\,\ref{spatial}.

We applied the GMM approach described in Sect.\,\ref{bimodal} to check whether the metallicity distribution of each of the defined regions is best represented by one or two Gaussian models. We found that a two-Gaussian distribution is preferred for all the cases. To estimate the mean metallicity and the associated uncertainty for each component and region, we followed the previously described method and employed MC simulations, building 1000 samples, randomly varying the metallicity of each star, and assuming Gaussian uncertainties. Figure\,\ref{spatial_gaussian} and Table\,\ref{met_spatial_table} show the obtained results. We measured a metallicity gradient with increasing metallicities towards the centre of the NSD that is present in both the metal-rich and metal-poor components. Figure\,\ref{spatial} shows the detected gradient. 

To check whether the metallicity gradients are real and to estimate their values, we employed MC simulations with 1000 iterations. In each iteration, we varied the metallicity of each point, assuming Gaussian uncertainties. We assumed a Galactic centre distance of 8.25\,kpc \citep{Gravity-Collaboration:2020aa} and computed the gradients for the negative and positive longitudes, separately for the metal-rich and metal-poor populations. We estimated the final gradient as the mean value of the gradients obtained for the positive and  negative longitudes. Our results indicate real gradients with increasing metallicity towards the centre of the NSD for both stellar populations. We ended up with values of $(3.4\pm1.7)\cdot10^{-4}$\,dex/pc for the metal-rich population and $(1.5\pm0.5)\cdot10^{-3}$\,dex/pc for the metal-poor population. These gradients are significant at approximately the 2$\sigma$ and 3$\sigma$ levels, respectively. However, the metal-poor trend might be influenced by the implicit assumption of the model and the higher metallicity uncertainties of the metal-poor stars \citep[see Fig.\,10 in ][]{Fritz:2020aa}. In any case, our result aligns with previous work finding a global metallicity gradient (without distinguishing between metal-rich and metal-poor stars) towards the innermost regions of the NSD \citep{Feldmeier-Krause:2022vm}.

We also found that the relative weight of the components does not change spatially within the uncertainties. The relative weight of the metal-rich component is slightly larger than the one obtained when combining all the stars without defining different regions at different latitudes. This is probably due to the metallicity gradient that shifted the mean metallicity of the metal-rich component towards lower metallicities when combining all the stars. Therefore, the metal-rich component encompasses more stars, and its relative weight is somewhat higher.

We also repeated the previous analysis on proper motions and $H-K_s$ colour for the metal-rich and metal-poor stellar populations in each region. We kept stars whose likelihood of belonging to either group is larger than 70\%, restricted the sample to targets whose proper motion uncertainty is below 1\,mas/yr, and limited the selection to stars with $[M/H]<0.6$\,dex, as previously explained. Figure\,\ref{spatial_proper} and Table\,\ref{met_spatial_table} show the results. In spite of the relatively low number of stars for the metal-poor population ($\sim 15-25$ stars for each spatial group), we found good agreement with the results obtained when analysing all the stars in Sect.\,\ref{bimodal}. Although there is not a significant difference between the mean proper motions for each stellar population (only in some cases we found a difference larger than 1\,$\sigma$), we detected that the velocity dispersion of the $\mu_b$ component of the metal-poor stellar population is $\gtrsim 2$\,$\sigma$ larger than that of the metal-rich stellar population in nearly all cases, except at $l = 0^\circ$, where it is $\sim 1$\,$\sigma$ larger. This supports the metal-poor component being kinematically hotter than the metal-rich one regardless of the spatial distribution. The bottom panel of Fig.\,\ref{spatial} shows the obtained results.

Finally, the colour analysis is also consistent with the previous results (Sect.\,\ref{extinction}). Thus, the difference between $H-K_s$ for the metal-rich and the metal-poor stellar population is always $\lesssim1\,\sigma$.

A similar analysis along the Galactic latitude, as presented in this section, is not feasible due to the low number of stars observed at higher NSD latitudes (see Fig.\,\ref{scheme}). This limitation is particularly true for metal-poor stars in the innermost fields.

\section{Discussion}

\subsection{Correlation between metallicity, velocity dispersion, and orbits}

We distinguished between metal-rich and metal-poor stars by applying a GMM approach and obtained that a two-Gaussian model best reproduces the data, as found in previous work \citep[e.g.][]{Schultheis:2021wf,Nogueras-Lara:2022tp,Nogueras-Lara:2023ab}. In principle, the presence of these two Gaussian components is just a mathematical description of the underlying metallicity distribution and is not necessarily related with the origin of metal-poor and metal-rich stars in the NSD. Actually, the metallicity distribution might be also explained by chemical evolution models \citep[e.g.][]{Grieco:2015ys,Friske:2023aa}, and thus the presence of metal-rich and metal-poor stars would be a consequence of the natural star-formation process in the NSD. However, the obtained correlation between metal-poor stars, high velocity dispersion, and chaotic/box orbits indicates a different origin for a significant fraction of metal-poor stars in comparison to the metal-rich ones. Our results suggest that the metal-poor stars are related to the innermost regions of the Galactic bar/bulge, which overlaps with the NSD. In this way, the obtained fraction of stars following chaotic orbits in Fig.\,\ref{orbits} aligns with the estimated $\sim20-30\%$ (decreasing for larger NSD radii) of contaminants from the Galactic bar/bulge reported in \citet{Sormani:2022wv}. Moreover, the metallicity distribution obtained for stars following chaotic orbits (likely associated with the Galactic bar/bulge) matches previous results for the innermost regions of the Galactic bar/bulge \citep[e.g.][]{Schultheis:2021wf}. Additionally, the higher velocity dispersion found for the metal-poor stellar population is in agreement with previous $\mu_b$ velocity dispersion measurements $\sim 2.5-3$\,mas/yr obtained for the Galactic bar/bulge in an inner Galactic region far away from any NSD contamination \citep[$l=0^\circ$ and $-4.5^\circ<b<-2.5^\circ$, e.g.][]{Kozowski:2006aa,Rattenbury:2007ab,Sanders:2019aa}.

Another possibility is that a fraction of the metal-poor stars might be remnants of dispersed, young gas-rich stellar clusters within a 1.5\,kpc radius of the Galactic centre, formed at z>4. These clusters have been suggested as precursors to the old metal-poor stars currently observed at the Galactic centre \citep[e.g.][]{van-Donkelaar:2023aa}. The accretion of stellar clusters, combined with in situ star formation, has been proposed as one of the formation channels of the nuclear star cluster at the centre of the Galaxy \citep[e.g.][]{tremaine75,Capuzzo-Dolcetta:1993aa,Antonini:2012aa,Antonini:2013ys,Perets:2014aa,Gnedin:2014fk,Arca-Sedda:2015aa,Tsatsi:2017aa,Arca-Sedda:2020ts}. Thus, \citet{van-Donkelaar:2023aa} used hydro-dynamical simulations to show that the infalling clusters can also be accreted by the NSD contributing to its growth.


We also found that the proper motion orientation of both metallicity components shows a different standard deviation, with the distribution of metal-poor stars being more scattered in comparison to the metal-rich component. This is compatible with the more chaotic motion that is expected under the influence of the Galactic bar/bulge. On the contrary, metal-rich stars are kinematically cooler and show orbits tracing the NSD \citep[e.g.][]{Schultheis:2021wf,Nieuwmunster:2024aa}.

The similar extinction observed for both the metal-poor and metal-rich components suggests a widespread distribution of both types of stars across the region. Hence, there is not a specific region along the line of sight in which either metal-rich or metal-poor stars are concentrated. This observation is also supported by the spatial analysis in Sect.\,\ref{extinction}, given that we found metal-rich and metal-poor stars distributed all over the field and with similar relative weights within the uncertainties.

\subsection{The NSD as a distinct kinematic structure}

Recently, \citet{Zoccali:2024aa} analysed the proper motion distribution of the NSD compared to control fields in the innermost region of the Galactic bar/bulge. Their findings did not reveal any clear, distinct kinematical signature of the NSD. This result challenges previous studies that identified the NSD as kinematically different from its surroundings based on line-of-sight velocities \citep{2015ApJ...812L..21S} and velocity dispersions \citep{Schultheis:2021wf}.

In this paper we limited our analysis to the NSD without any comparison with the surrounding region. Nevertheless, we found the presence of two kinematically distinct components with different metallicities and orbits. We detected that the metal-poor component is dominated by chaotic/box orbits that are likely associated with the Galactic bar and have a metallicity distribution also in agreement with previous work on the inner bar/bulge \citep[carried out in fields without contamination from the NSD, e.g.][]{Schultheis:2021wf}. On the other hand, the dominant metal-rich component is kinematically cooler, given its lower velocity dispersion, and undergoes z-tube orbits, better tracing the NSD \citep[e.g.][]{Nieuwmunster:2024aa}. The fact that we observed these two different stellar populations with different orbits and kinematics — with one more closely associated with the Galactic bar/bulge— suggests that the NSD is a different kinematic structure, and this agrees with previous work \citep[][]{2015ApJ...812L..21S,Schultheis:2021wf}, showing that it is possible to kinematically distinguish the NSD. Additionally, having a metal-rich and kinematically cool NSD different from the surrounding Galactic bar/bulge also agrees with the results of previous works on external Milky Way-like galaxies, where these structures have been identified via their different kinematics and metallicities \citep[e.g.][]{Gadotti:2019aa,Bittner:2020aa,Gadotti:2020aa}.

We would like to emphasise that \citet{Zoccali:2024aa} do not exclude the possibility of a kinematically different NSD in comparison to the surrounding Galactic bar/bulge. They only state that the region corresponding to the NSD is dominated by stars moving eastwards, which is compatible with the CMZ hiding the counter-rotating stars, but this does not mean that the NSD is kinematically different. Nevertheless, analysing the proper motion component parallel to the Galactic plane is very challenging due to the extreme extinction that dominates this region. An in-detail analysis of velocity dispersions, especially the proper motion component perpendicular to the Galactic plane, including fields from the innermost Galactic bar/bulge, is likely to confirm that the NSD is kinematically distinct.

\subsection{The metallicity gradient and the NSD formation}

Stellar population analyses along and across the line of sight towards the NSD have suggested the presence of radial age and metallicity  gradients towards the centre of the NSD \citep[e.g.][]{Nogueras-Lara:2019ad,Nogueras-Lara:2022ua,Feldmeier-Krause:2022vm,Nogueras-Lara:2023aa,Nogueras-Lara:2023ab,Nogueras-Lara:2024aa}. In this way, the innermost regions of the NSD contain a predominantly old ($\gtrsim 7$\,Gyr) and metal-rich stellar population, while a significant intermediate-age stellar population ($\sim$40\% of the stellar mass with ages between 2-7\,Gyr) is found towards the outer edge of the NSD, where the metallicity is lower. In contrast, \citet{Sanders:2023aa} analysed the period-age relation of Mira variables in the NSD and found only a weak indication of inside-out formation. They attributed this to potential dynamical mixing and/or the smoothing effect of their spline model, which might hide a significant population of Mira variables with ages around $5-6$\,Gyr (periods $\sim400$,days). These variables could potentially trace the gradient, but they have large uncertainties.

We observed that the metal-rich and metal-poor components of the NSD are uniformly distributed across the field, and both exhibit a metallicity gradient with a higher metallicity towards the centre of the NSD. These metallicity gradients correspond to an increase of $(3.4\pm1.7)\cdot10^{-4}$\,dex/pc and $(1.5\pm0.5)\cdot10^{-3}$\,dex/pc for the metal-rich and metal-poor stellar populations, respectively. To compare our results with the metallicity gradient obtained by \citet{Feldmeier-Krause:2022vm}, which does not distinguish between metal-rich and metal-poor stars, we estimated the metallicity increase per parsec using the metallicity values at the edge and centre of a region located $\pm19$\,pc in latitude from the centre of the NSD, as specified in their Fig.\,11. We obtained a metallicity gradient of approximately $5 \times 10^{-4}$\,dex/pc, which roughly corresponds to the mean of the gradients we found for the metal-rich and metal-poor stars.

The detected gradients align with the currently accepted inside-out formation scenario of the NSD through bar-driven mechanisms \citep[e.g.][]{Bittner:2020aa,Nogueras-Lara:2023aa}. Similar gradients have been found in Milky Way-like galaxies analysed in the TIMER survey \citep[e.g.][]{Gadotti:2019aa,Bittner:2020aa} and can be attributed to the growth of the galaxy bar that funnels gas from different galactic radii towards the centre as it evolves (the more internal, the more metal rich). Furthermore, in evolved systems, it is anticipated that metallicity will rise towards the centre due to the enhanced difficulty for stellar feedback processes to expel gas as it becomes more tightly bound within the galaxy's potential well. This is compounded by the potential rapid increase in metallicity in star-forming systems.

Interestingly, the metallicity gradients that we found when analysing different NSD lines of sight align well with the previous results from \citet{Nogueras-Lara:2023ab}, who also detected metallicity gradients for metal-rich and metal-poor stars along the line of sight towards the nuclear star cluster (which also contains stars from the NSD that dominate the stellar population in front of the nuclear star cluster). These similarities may suggest a common formation scenario for both the nuclear star cluster and the disc, as discussed by \citet{Nogueras-Lara:2023ab}.

Additionally, the metal-rich stellar population follows a rotation pattern similar to that of the gas in the CMZ \citep{Schultheis:2021wf}. Thus, it is likely that the metal-rich stars formed in situ from gas funnelled towards the Galactic centre via bar-driven processes, contributing to the inside-out growth of the NSD. This also justifies why these stars undergo z-tube orbits and have a lower velocity dispersion in comparison to the metal-poor component. Conversely, the metallicity gradient observed for metal-poor stars, which are kinematically hotter and follow Galactic bar-related orbits, may be caused by the combination of Galactic bar interlopers and/or stars accreted from clusters within a radius of 1.5\,kpc at $z > 4$ \citep{van-Donkelaar:2023aa}.

\section{Conclusion}

In this paper, we have investigated the proper motion and orbit distribution of stars with different metallicities in the NSD. We obtained that metal-rich stars show a lower velocity dispersion in comparison to metal-poor ones and that they are dominated by stars following z-tube orbits that are representative of the NSD \citep{Nieuwmunster:2024aa}. Conversely, metal-poor stars are kinematically hotter, and a significant fraction of them undergo chaotic/bar orbits that are likely related to the Galactic bar. We also found that the proper motion orientation of metal-rich stars shows a sharper profile parallel to the Galactic plane in comparison with stars that are metal poor.

We used the statistical correlation between distance and extinction in the NSD \citep{Nogueras-Lara:2022aa} to explore whether stars with different metallicities show a special location along the line of sight. We concluded that the average near-infrared colour of both stellar populations is similar within the uncertainties, and there is no significant difference in extinction. Therefore, we did not detect any preferred location for stars with different metallicities, and they seem to be present all over the observed regions.

Finally, we analysed the metallicity distribution of stars at different longitudes and found that metal-rich and metal-poor stars are present throughout the NSD. We also detected the presence of metallicity gradients for both stellar populations. These gradients have increasing metallicities towards the central regions of the NSD. Our results support an inside-out formation of the nuclear stellar disc, as inferred from studies of external Milky Way-like galaxies \citep[e.g.][]{Gadotti:2019aa,Bittner:2020aa}. We interpret the gradient in the metal-rich population as being a consequence of in situ formation from gas funnelled towards the Galactic centre through bar-driven mechanisms \citep[e.g.][]{Sormani:2019aa}. This justifies the dominant z-tube orbits and the lower velocity dispersion and kinematics that align with the rotation velocity of gas from the CMZ \citep{Schultheis:2021wf}.

\begin{acknowledgements}
  
This work is based on observations made with ESO Telescopes at the La Silla Paranal Observatory under program ID 0101.B-0354. FN-L gratefully acknowledges the sponsorship provided by the European Southern Observatory through a research fellowship. FF is supported by a UKRI Future Leaders Fellowship (grant no. MR/X033740/1). MCS acknowledges financial support from the European Research Council under the ERC Starting Grant ``GalFlow'' (grant 101116226). BT acknowledges the financial support from the Wenner-Gren Foundation (WGF2022-0041).

\end{acknowledgements}

\bibliography{BibGC.bib}

\onecolumn

\appendix

\section{Parameters of the spatial analysis of the two-Gaussian metallicity distribution}

\begin{figure*}[h]
\centering
                 
   \includegraphics[width=0.9\linewidth]{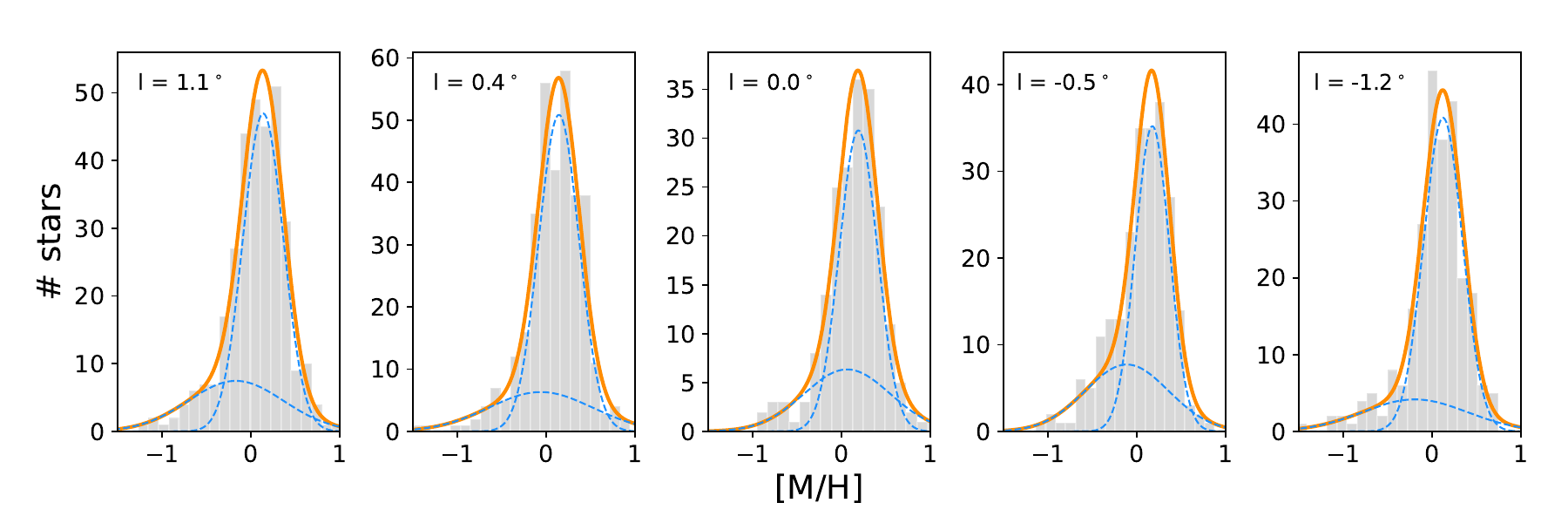}
   \caption{Two-Gaussian decomposition of the NSD metallicity distribution for different longitudes. The orange line shows the result of the GMM analysis, whereas the blue dashed lines depict each of the Gaussian components.}

\label{spatial_gaussian}
\end{figure*}

\begin{table*}[h]

\caption{Characterisation of the two-Gaussian decomposition of the NSD metallicity distribution at different longitudes.}
\label{met_spatial_table} 
\begin{center}

\begin{tabular}{c|ccc|cccc|cc}
\multicolumn{1}{c}{} &  &  & \multicolumn{1}{c}{} &  &  &  & \multicolumn{1}{c}{} & & \tabularnewline
\hline 
\hline 
$l$ & $W$ & $[M/H]$ & $\sigma_{[M/H]}$ & $\mu_l$ & $\sigma_{\mu_l}$ & $\mu_b$ & $\sigma_{\mu_b}$ & $H-K_s$ & $\sigma_{H-K_s}$\tabularnewline
\hline 
1.1 & {0.72 $\pm$ 0.05} & {0.15 $\pm$ 0.02} & {0.25 $\pm$ 0.01} & {-1.74 $\pm$ 0.18} & {2.50 $\pm$ 0.13} & {0.19 $\pm$ 0.14} & {1.90 $\pm$ 0.13} & {2.58 $\pm$ 0.05} & {0.64 $\pm$ 0.03}\tabularnewline
 & {0.28 $\pm$ 0.05} & {-0.17 $\pm$ 0.07} & {0.51 $\pm$ 0.05} & {-2.36 $\pm$ 0.63} & {3.04 $\pm$ 0.44} & {-0.20 $\pm$ 0.73} & {3.44 $\pm$ 0.57} & {2.42 $\pm$ 0.15} & {0.70 $\pm$ 0.08}\tabularnewline
\hline 
0.4 & {0.66 $\pm$ 0.06} & {0.16 $\pm$ 0.02} & {0.25 $\pm$ 0.01} & {-1.55 $\pm$ 0.18} & {2.65 $\pm$ 0.10} & {0.34 $\pm$ 0.11} & {1.66 $\pm$ 0.11} & {2.43 $\pm$ 0.03} & {0.49 $\pm$ 0.02}\tabularnewline
 & {0.34 $\pm$ 0.06} & {-0.03 $\pm$ 0.05} & {0.50 $\pm$ 0.06} & {-3.97 $\pm$ 0.89} & {3.29 $\pm$ 0.42} & {1.71 $\pm$ 0.69} & {2.55 $\pm$ 0.50} & {2.43 $\pm$ 0.17} & {0.62 $\pm$ 0.09}\tabularnewline
\hline 
0.0 & {0.64 $\pm$ 0.09} & {0.20 $\pm$ 0.04} & {0.26 $\pm$ 0.03} & {-0.81 $\pm$ 0.25} & {2.62 $\pm$ 0.13} & {0.06 $\pm$ 0.16} & {1.67 $\pm$ 0.12} & {2.33 $\pm$ 0.04} & {0.41 $\pm$ 0.03}\tabularnewline
 & {0.36 $\pm$ 0.09} & {0.05 $\pm$ 0.06} & {0.50 $\pm$ 0.09} & {-0.74 $\pm$ 0.72} & {2.38 $\pm$ 0.35} & {-0.50 $\pm$ 0.68} & {2.18 $\pm$ 0.51} & {2.26 $\pm$ 0.16} & {0.50 $\pm$ 0.09}\tabularnewline
\hline 
-0.5 & {0.70 $\pm$ 0.05} & {0.17 $\pm$ 0.03} & {0.25 $\pm$ 0.02} & {-2.02 $\pm$ 0.24} & {2.74 $\pm$ 0.14} & {0.26 $\pm$ 0.14} & {1.55 $\pm$ 0.12} & {2.52 $\pm$ 0.05} & {0.55 $\pm$ 0.04}\tabularnewline
 & {0.30 $\pm$ 0.05} & {-0.18 $\pm$ 0.10} & {0.48 $\pm$ 0.08} & {-1.70 $\pm$ 0.51} & {2.36 $\pm$ 0.49} & {-0.60 $\pm$ 0.58} & {2.63 $\pm$ 0.47} & {2.54 $\pm$ 0.10} & {0.45 $\pm$ 0.07}\tabularnewline
\hline 
-1.2 & {0.77 $\pm$ 0.05} & {0.13 $\pm$ 0.01} & {0.25 $\pm$ 0.01} & {-0.45 $\pm$ 0.21} & {2.61 $\pm$ 0.15} & {0.08 $\pm$ 0.13} & {1.57 $\pm$ 0.12} & {1.95 $\pm$ 0.04} & {0.48 $\pm$ 0.04}\tabularnewline
 & {0.23 $\pm$ 0.05} & {-0.23 $\pm$ 0.12} & {0.57 $\pm$ 0.06} & {-0.80 $\pm$ 0.51} & {1.79 $\pm$ 0.41} & {-0.01 $\pm$ 0.62} & {2.21 $\pm$ 0.33} & {2.21 $\pm$ 0.17} & {0.59 $\pm$ 0.11}\tabularnewline
\hline 
\multicolumn{1}{c}{} &  &  & \multicolumn{1}{c}{} &  &  &  & \multicolumn{1}{c}{} & & \tabularnewline
\end{tabular}

\end{center}
\footnotesize
\textbf{Notes.} $W$, $[M/H]$, and $\sigma_{[M/H]}$ correspond to the results from the GMM analysis of the two-Gaussian distribution, where $W$ indicates the relative weight of each of the Gaussian components. $\mu_l$, $\sigma_{\mu_l}$, $\mu_b$, and $\sigma_{\mu_b}$ are in mas/yr. $H-K_s$ and $\sigma_{H-K_s}$ are in Vega magnitudes.

\end{table*}

\begin{figure*}[h]
   \centering
   \includegraphics[width=0.9\linewidth]{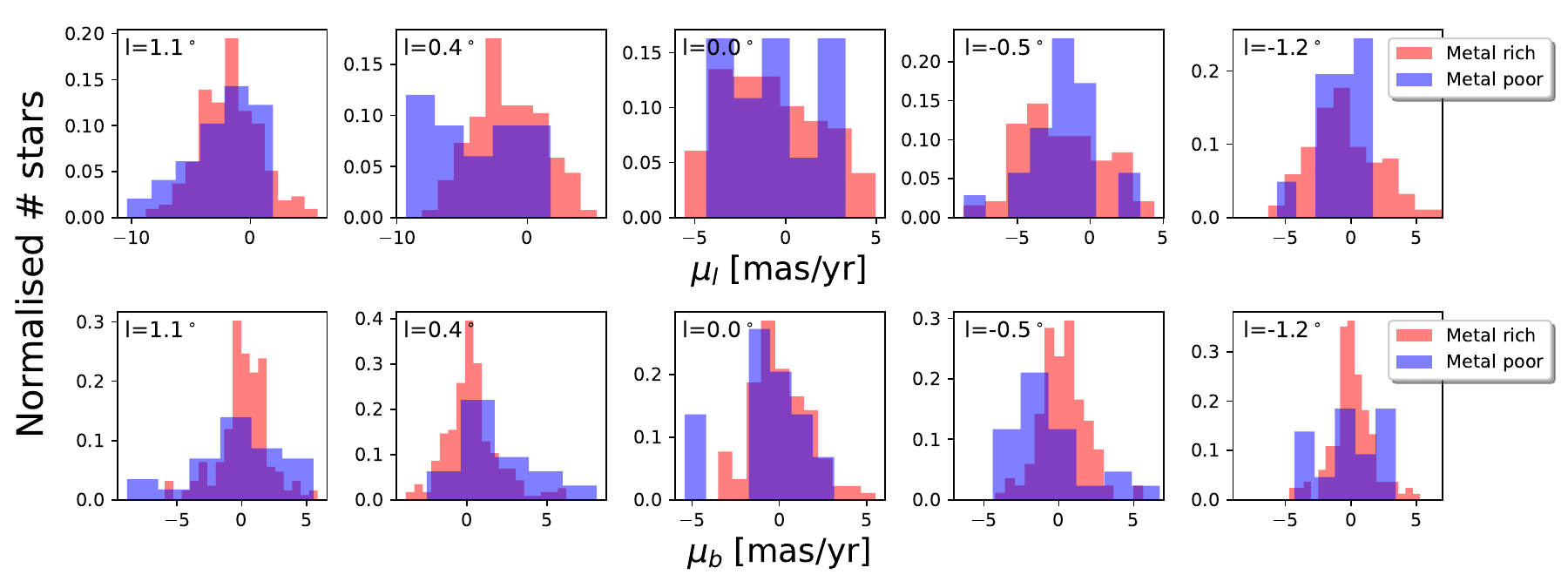}
   \caption{Proper motion distribution of metal-rich and metal-poor stars in the NSD for different longitudes following the GMM analysis in Sect.\,\ref{spat}.}

\label{spatial_proper}
\end{figure*}

\end{document}